\documentclass[%
 reprint,
 amsmath,amssymb,
 aps,
]{revtex4-2}

\usepackage{graphicx}
\usepackage{dcolumn}
\usepackage{bm}
\usepackage{color}

\begin{document}

\preprint{APS/123-QED}

\title{Crossover from Anomalous to Normal Diffusion: Ising Model with Stochastic Resetting}

\author{Yashan Chen $^1$ $^2$}
\author{Wei Zhong $^1$}
 \email{zhongwei2284@hotmail.com}

\affiliation{%
 $^1$ MinJiang Collaborative Center for Theoretical Physics, College of Physics and Electronic Information Engineering, Minjiang University, Fuzhou 350108, P. R. China.\\
 $^2$ Fujian Yongrong Jinjiang Co.,Ltd., Fuzhou, P. R. China.
}%



\date{\today}

\begin{abstract}
In this paper, we investigate the dynamics of the two-dimensional Ising model with stochastic resetting, utilizing a constant resetting rate procedure with zero-strength initial magnetization. Our results reveal the presence of a characteristic rate $r_c \sim L^{-z}$, where $L$ represents the system size and $z$ denotes the dynamical exponent. Below $r_c$, both the equilibrium and dynamical properties remain unchanged. At the same time, for $r > r_c$, the resetting process induces a transition in the probability distribution of the magnetization from a double-peak distribution to a three-peak distribution, ultimately culminating in a single-peak exponential decay. Besides, we also find that at the critical points, as $r$ increases, the diffusion of the magnetization changes from anomalous to normal, and the correlation time shifts from being dependent on $L$ to being $r$-dependent only.
\end{abstract}

\maketitle


\section{Introduction \label{sce1}}

Stochastic resetting, a stochastic process that returns to its initial state, has garnered significant attention in the past decade \cite{evans2020stochastic,montero2017continuous,gupta2022stochastic}. 
﻿

One of its key characteristics in systems with stochastic resetting is the emergence of resetting-induced non-equilibrium stationary states (NESS), as first reported by Evans and Majumdar \cite{evans2011diffusion} in the context of one-dimensional Brownian motion. They demonstrated that by repeatedly returning the Brownian particle to its initial position at a constant resetting rate, a nonequilibrium stationary state would arise. Subsequently, this resetting-induced NESS has been observed in other diffusion processes, including multi-dimensional diffusion \cite{evans2014diffusion}, coagulation-diffusion processes \cite{durang2014statistical}, underdamped Brownian motion \cite{gupta2019stochastic}, active particle systems \cite{evans2018run,kumar2020active,sar2023resetting}, and quantum many-body systems \cite{perfetto2021designing,turkeshi2022entanglement}. 
﻿

Moreover, the interplay between diffusion under resetting and critical phenomena has been investigated \cite{singh2020resetting,kusmierz2014first,campos2015phase}. However, the impact of stochastic resetting on the critical behavior of a system remains unclear. Recently, Magoni, Majumdar, and Schehr \cite{magoni2020ising} discussed the Glauber Ising model with stochastic resetting under a constant resetting rate and observed a NESS regime in the $r-T$ phase diagram, where $r$ represents the resetting rate and $T$ denotes the temperature. Notably, at the critical point, the influence of stochastic resetting on the dynamics of the system was not addressed.

In a recent paper, \cite{zhong2018generalized}, it was reported that at the critical point of the Ising model, the magnetization undergoes anomalous diffusion, a phenomenon that can be described by fractal Brownian motion (fBm) \cite{panja2010generalized}. Anomalous diffusion, characterized by a non-linear mean-square deviation over time, has been observed in diverse systems such as financial markets \cite{mantegna1999introduction}, bacterial systems \cite{ariel2015swarming}, and other disordered media \cite{bouchaud1990anomalous}. More recently, several research groups have demonstrated that stochastic resetting can alter the diffusion regimes for certain types of anomalous diffusion, including continuous time random walk (CTRW) \cite{kusmierz2019subdiffusive,masoliver2019anomalous,maso2019anomalous} and scaled Brownian motion (SBM) \cite{bodrova2019nonrenewal,bodrova2019sbm}, within comb-like structures \cite{singh2021backbone,antonio2020comb}, and for systems exhibiting anomalous diffusion of the fBm type \cite{wang2021time}. Hence, it is natural to inquire whether stochastic resetting can similarly modify the diffusion regime in the context of the Ising model.
﻿

In this study, we investigate the two-dimensional Ising model on an $L \times L$ square lattice at the critical point with stochastic resetting. We employ a constant resetting rate protocol, whereby the system can reset to its initial state (with the initial magnetization set to be $0$) at each step with a fixed rate $r$. Initially, we compute the probability distribution of the order parameter $P(m)$ for various resetting rates. Subsequently, we explore the impact of stochastic resetting on the system's dynamics by measuring the mean-square deviation and the correlation length of the order parameter.

Our findings reveal a characteristic rate $r_c \sim L^{-z}$, where $z$ represents the dynamical exponent. Below $r_c$, the probability distribution $P(m)$ does not deviate from the pure Ising model, displaying double-peak distributions, and stochastic resetting does not alter the system's dynamics, with the correlation length continuing to scale as $ \sim L^z$. Conversely, for $r > r_c$, as $r$ increases, $P(m)$ transitions from a double-peak distribution to a three-peak distribution, ultimately culminating in a single-peak exponential decay. Additionally, we observe a crossover from anomalous to normal diffusion in the mean-square deviation of the order parameter as $r$ increases and the correlation time shifts from being dependent on $L$ to solely dependent on $r$. These results serve as a valuable reference for investigating how stochastic resetting can impact systems with subdiffusion characteristics of the fractal Brownian motion type.

The paper is structured as follows: In Section \ref{sec2}, we present the Ising model with stochastic resetting, providing a detailed explanation of the constant resetting rate protocol. Then, in Section \ref{sec3}, we examine how the stochastic resetting process influences the diffusion of the order parameter at the critical points. Finally, the paper concludes in Section \ref{sec4}.
 
\section{Stochastic resetting on the Ising model\label{sec2}}

We first briefly introduce the two-dimensional ($2D$) Ising model. Its Hamiltonian is described as
 
 \begin{equation}
{\cal H}=-J\sum_{\langle ij\rangle} s_i s_j, 
 \end{equation}
where the coupling strength $J$ is set to be $1$, and the system lays upon an $L\times L$ square lattice with periodic boundary conditions. At each site $i$, the spin value can be $s_i=\pm 1$. $\langle ij\rangle$ denotes the summation runs over all nearest neighbors.
 
The system is simulated with the Metropolis algorithm, i.e., at each time step, a single spin is selected to do the flip attempt. The possible energy change $\Delta E$ is measured. If $\Delta E\leq 0$, the flip is accepted with unit probability; otherwise, the flip is accepted via the Metropolis probability $e^{-\Delta E/(k_B T)}$, where $k_B$ is the Boltzmann constant and $T$ is the temperature. A sweep (or a unit of time) refers to one flip attempt per spin. 

Throughout the simulation, we employ stochastic resetting in the system. In the literature, various protocols of stochastic resetting have been discussed. One of the simplest protocols is the constant resetting rate protocol, where the system is reset to its initial state with a consistent resetting rate. However, this method is unsuitable for memoryless resetting processes \cite{bodrova2019nonrenewal}. Therefore, several alternative protocols have been proposed, including position-dependent resetting rates \cite{evans2011diffusion2}, time-dependent resetting rates \cite{pal2016diffusion,shkilev2017continuous}, and resetting rates distributed as a power-law in waiting times \cite{nagar2016diffusion}. 

\begin{figure}[htb]
  \includegraphics[width=0.95 \linewidth]{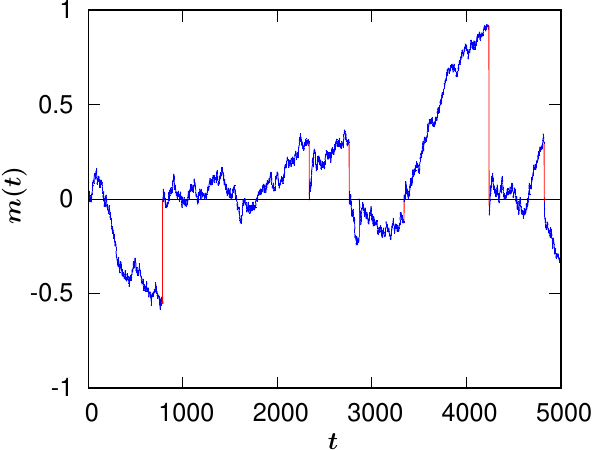}
  \hspace{3mm}  \caption{The stochastic resetting process in the Ising model for $L=64$ and $T=2.0$. The magnetization resets to its initial value $m_0=0.0$ with a constant resetting rate $r=0.001$. The red vertical lines represent the resetting cases observed within $t\leq 5000$. }
  \label{fig_snap}
\end{figure}

For our study, we utilize the constant resetting rate protocol (as depicted in Fig. \ref{fig_snap}). This means that at each time step (Monte Carlo sweep), a random number within the range $[0, 1)$ is generated. If the random number is less than our chosen resetting rate $r$, then the system resets to its initial state with $m \rightarrow m_0$, where we choose $m_0 = 0$; otherwise, the system continues to evolve using the Metropolis algorithm. Note that the probability of the resetting event to occur at a given time is $r dt$, where $dt$ is the time step. In our simulations, $dt=1$ is employed, and then, $r$ becomes the probability of the resetting event occurring at a given time. 
 
\begin{figure*}[htb]
   \includegraphics[width=0.52  \linewidth]{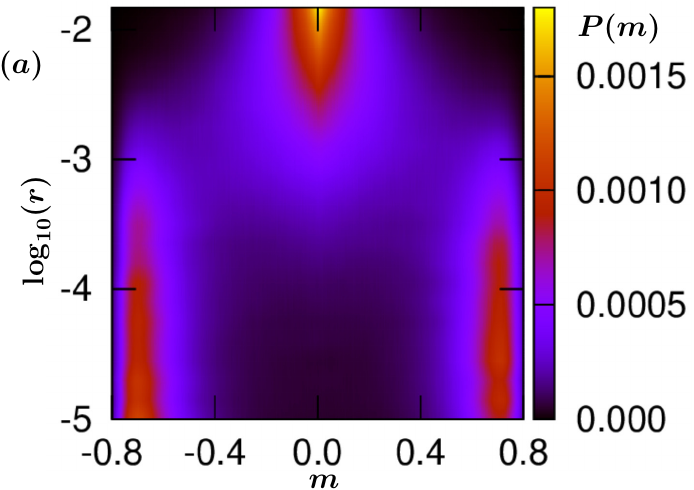} \hspace{1mm}          
  \includegraphics[width=0.44 \linewidth]{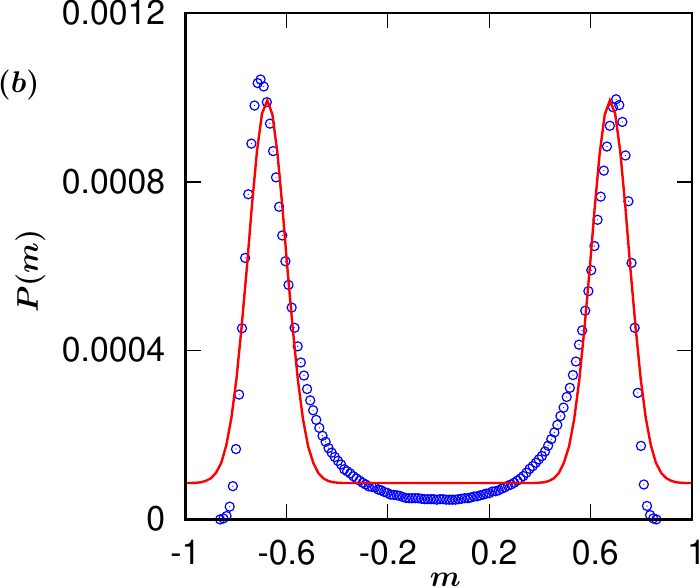}\\ 
  \includegraphics[width=0.46\linewidth]{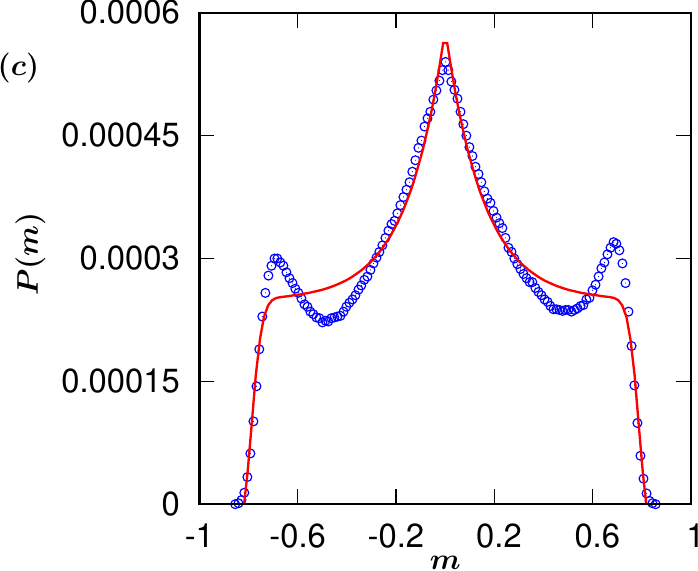} \hspace{5mm}          
   \includegraphics[width=0.46\linewidth]{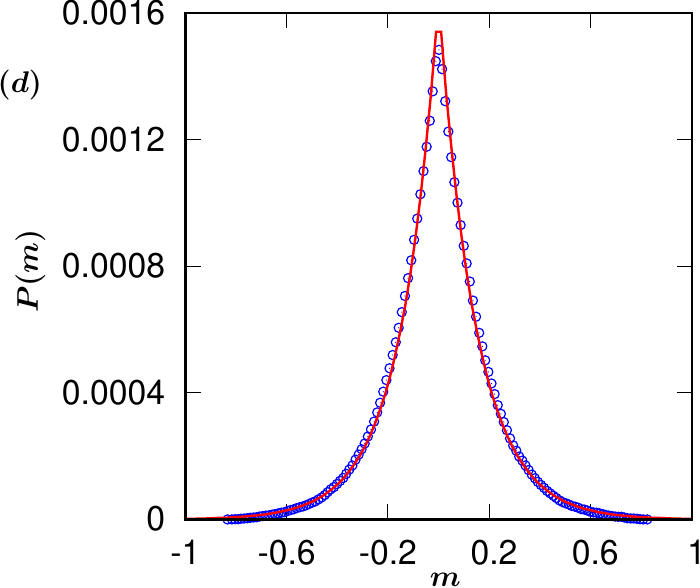} 
  
  \caption{(a) The probability distributions of the magnetization $P(m)$ for $L=64$ with different resetting rate at the critical temperature. It shows that when the resetting rate $r$ grows from $0$ to $1$, $P(m)$ varies from a double-peak distribution to a third-peak one. When $r\gtrsim 1/L$, the probability distribution function reduces to the exponential distribution. (b)-(d) The probability distributions of the magnetization for $L=64$ with resetting rate $r=0.000005$, $0.00128$ and $0.01024$, respectively. The red solid lines denote the plot fit via (b) Eq. (\ref{prob_1}), (d) Eq. (\ref{prob_2}), and (c) their combinations.
  }
  \label{fig_prob}
\end{figure*}

\section{Crossover from anomalous to normal diffusion \label{sec3}} 

For a system that evolves with a constant stochastic resetting rate $r$, typically, the resetting event disrupts the deterministic progression of a system. As a result, the system's state at time $t$ depends solely on the time $\tau$ elapsed since the last resetting event \cite{evans2011diffusion,magoni2020ising}. 

Specifically, in the context of a Poisson process with a rate $r$, the probability of no reset occurring in the time interval $[t - \tau, t]$ is denoted by $e^{-r\tau}$, while the probability of a reset occurring in the infinitesimal time interval $d\tau$ is denoted by $rd\tau$. Consequently, the probability distribution of the time elapsed since the last resetting event being $\tau$ is succinctly described by $p(\tau)d\tau = re^{-r\tau}$.

 Naturally, there's the possibility that the system progresses until time $t$ without any reset, associated with the probability $e^{-rt}$. This probability signifies the likelihood of no reset happening throughout the whole time interval $[0, t]$. Therefore, the distribution of a random variable ${\cal O}$ under stochastic resetting goes as \cite{evans2011diffusion,magoni2020ising}  
\begin{equation}
P_r({\cal O},t)=r\int_0^t d\tau e^{-r \tau} P({\cal O},\tau)+e^{-rt} P({\cal O},t),
\label{eq_2}
\end{equation}
where the first term on the right-hand side (RHS) represents the probability distribution contributed by the occurrence of a reset process at time $t-\tau$ for any possible $\tau\in (0,t)$, where $\tau$ denotes the time duration between the last resetting event and $t$. The second term on the RHS is the contribution from those events when no resetting occurs within the time interval [0,t]. When $t$ is large, $e^{-rt}$ approaches $0$.  Therefore, we simplify Eq. (\ref{eq_2}) as
\begin{equation}
P_r({\cal O},t)=r\int_0^t d\tau e^{-r \tau} P({\cal O},\tau).
\label{prob}
 \end{equation}
which means to obtain the probability distribution of the random variable ${\cal O}$, we only need to know the information of $P_0({\cal O},\tau)$.

\subsection{Probability distribution of the order parameter at the critical temperature \label{sec3.1}}

For the Ising model, we select the order parameter $m=\frac{1}{N} \sum_{i=0}^N s_i$ as the target random variable. Magoni and colleagues \cite{magoni2020ising} have explored the probability distribution of the order parameter for $1D$ and $2D$ Ising models. They argued that the critical temperature $Tc$ remains unchanged for different resetting rates $r$. Besides, the resetting process gives rise to a 'pseudoferro' phase, {\it i.e.,} a resetting-induced non-equilibrium stationary state (NESS), for $r > r^{\star}(T)$ and $T > Tc$, where $r^{\star}(T)$ represents a threshold value of the resetting rate. However, at the critical point, the analytical description of $P(m)$ is lacking. 

Without resetting, Binder suggested that $P(m)$ should have a double-peak behavior as follows \cite{binder1982monte}
 
\begin{multline}
P(m) = \frac{L^{d/2}}{2 \pi k_B T \chi^{L}} \frac{1}{2} \exp\left( \frac{(m-m^{L})^2 L^d}{2 k_B T \chi^{L}}\right) \\
 + \frac{L^{d/2}}{2 \pi k_B T \chi^{L}} \frac{1}{2} \exp\left( \frac{(m+m^L)^2 L^d}{2 k_B T \chi^{L}}\right)  
 \label{prob_1}
\end{multline}
where $d$, $\chi^L$, and $m^L$ represent the space dimension of the system, the susceptibility, and the peak value of the magnetization, respectively. As an example, in Fig. \ref{fig_prob} (b), we obtain the best fit of Eq. (\ref{prob_1}) to the simulation results for $L=64$ at the critical point with tiny resetting rate.

We also know that for a time sequence of the order parameter, the correlation time behaves as $\tau \sim \xi^z$, where $\xi$ is the correlation length and $z\approx 2.1665$ \cite{nightingale1996dynamic} is the dynamical exponent. Therefore, with the introduction of stochastic resetting to the system, to maintain the critical behavior of the Ising model, we can postulate that the inverse value of the resetting rate should exceed the correlation time, i.e., $1/r \gtrsim \tau$, and $\tau\sim \xi^z$. At $Tc$, the correlation length $\xi$ approaches $L$, thus we have $r\lesssim L^{-z}$. In this range of resetting rates, the probability distribution of the order parameter resembles $P (m)$ at $r = 0$.

 \begin{figure}[htbp]
  \includegraphics[width=0.88 \linewidth]{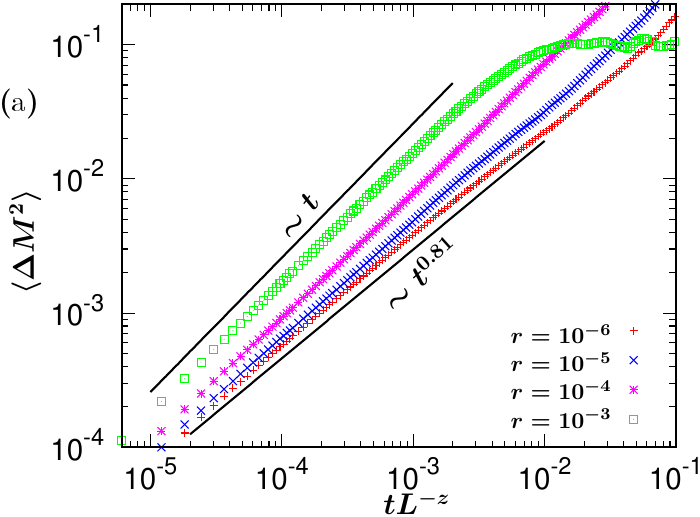}\\
  \includegraphics[width=0.96 \linewidth]{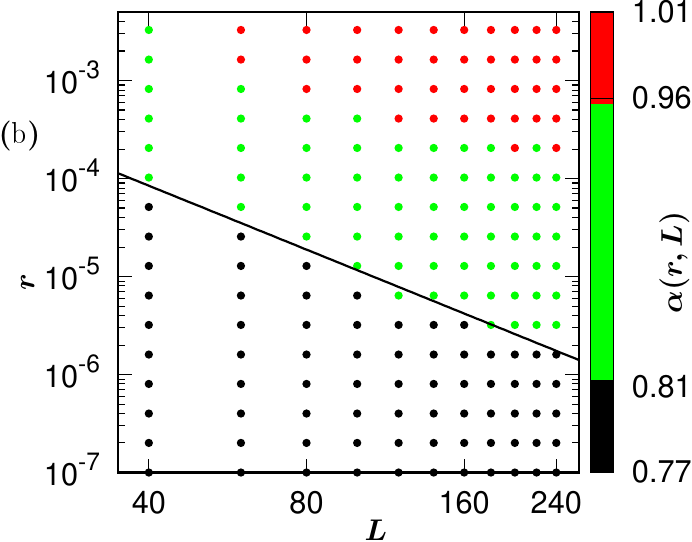}
  \hspace{3mm}  \caption{(a) The MSD of the magnetization for the Ising model with different resetting rates $r$. The system size is set to be $L=256$. The figure shows that the magnetization experiences subdiffusion with a low resetting rate, and normal diffusion with a high resetting rate. Especially, when $r\lesssim L^{-z}$, the MSD of the magnetization behaviors as $\sim t^{\gamma /(\nu z)}$, with $\gamma=1.75$, $\nu=1.0$ are equilibrium exponents, and $z\approx 2.1665$ is the dynamical exponent. (b) The anomalous exponents $\alpha(r, L)$ of the magnetization for different system sizes and resetting rates. The solid line represents $r_c\sim L^{-z}$. Below this solid line, the stochastic resetting does not affect the dynamics of the system, {\it i.e.}, $\alpha(r, L)\approx \alpha_c$. }
  \label{fig_msd}
\end{figure}

 For a large resetting rate $r \gtrsim 1/L$, the resetting is so potent that it ensnares the system in a profound non-equilibrium state, and the long-time memory of the magnetization is elapsed. Then, the distribution $P(m,\tau)$ behavior as a Gaussian distribution
\begin{equation}
P(m,\tau) \sim e^{-(m-m_0)^2/(4 D \tau)}/(4\pi D \tau),
\label{gaussian}
\end{equation}
with $D$ representing a diffusion constant.

Substituting Eq. \ref{gaussian} into Eq. \ref{prob}, we have 
\begin{equation}
P_r(m)  \sim r\int_0^t d\tau e^{-r \tau} e^{-(m-m_0)^2/(4 D \tau)}/(4\pi D \tau),
\end{equation}
which lead to 
\begin{equation}
P_r(m) \sim \sqrt{r/D} e^{-\sqrt{r/D} |m-m_0|},
\label{prob_2}
\end{equation}
resulting in a probability distribution with exponential decay as shown in Fig. (\ref{fig_prob}) (d).  

Between those two extremes (small and large resetting rate), the interplay between fluctuation and resetting leads to an intermediate three-peak distribution with the formula is a combination of Eq. (\ref{prob_1}) and Eq. (\ref{prob_2}). The simulation results depicted in Fig. \ref{fig_prob} validate these processes. It's important to note that the alterations in the shapes of the distribution functions signify the breakdown of symmetries. This indicates that the dynamics of the system, particularly the diffusion behavior, could be altered \cite{bouchaud1990anomalous}.
 
It has already been discussed that the stochastic resetting process can modify the diffusion process, resulting in changing the diffusion type from anomalous to normal, or subdiffusion to superdiffusion in different systems \cite{maso2019anomalous,antonio2020comb,bodrova2019nonrenewal}. Besides, recent studies \cite{zhong2018generalized, walter2015introduction} imply that the Ising model at the critical point provides a good reference to explore anomalous diffusion that belongs to fractal Brownian motion (fBm). Therefore, in the next subsection, we focus on the diffusion of the order parameter in the Ising model at $Tc$. 

\subsection{Diffusion of the order parameter \label{sec3.2}}

To quantitatively depict the diffusion of the order parameter, we define the mean-square deviation (MSD) as

\begin{equation}
\langle \Delta m^2\rangle=\langle [m(t)-m(0)]^2\rangle.
\end{equation}
 
Ref. \cite{zhong2018generalized} showed that at $T_c$, the MSD of the order parameter, without the resetting process, behavior as
\begin{equation}
\langle \Delta m^2\rangle = L^d t^\alpha,
\end{equation}
where $d=2$ is the spatial dimension, and the anomalous exponent $\alpha=)D-d+\gamma/\nu) / z$, with $d$ is the spatial dimension of the system and $D$ is the tagged dimension. For example, if we focus on a tagged line, then $D=1$. $\gamma=1.75$ and $\nu=1$ are two equilibrium exponents. 

For the $2D$ Ising model, when we focus on the bulk magnetization of the system, we have $D=d=1$, then $\alpha= \gamma/(\nu z)\approx 0.81$. It means that the magnetization of the $2D$ Ising model experiences subdiffusion at the critical point. 

Further, Ref. \cite{zhong2018generalized} reported that anomalous diffusion is popular in the Ising-like systems at the critical point. By analyzing the autocorrelation function of the restoring force, which is the force that resists the change of the magnetization, Ref. \cite{zhong2018generalized} confirmed that the observed subdiffusion of the magnetization probability belongs to the fractal Brownian motion type \cite{panja2010generalized}. We also measure the force autocorrelation in the Appendix and our results support the argument of Ref. \cite{zhong2018generalized}.

For the Ising model with stochastic resetting, Fig. \ref{fig_msd} (a) verifies that at a small resetting rate, the behavior of $\langle \Delta m^2\rangle$ is identical to the situation without resetting. When $r$ increases, the anomalous diffusion of $m$ is still observed. However, the anomalous exponent increases. For very large $r\gtrsim 1/L$, $\alpha\approx 1$ is obtained. These results reveal that the stochastic resetting induces a crossover behavior, {\it i.e.}, the order parameter experiences a crossover from anomalous to normal diffusion with increasing $r$. 

Note that for large times, the MSD of the magnetization will saturate to a value dependent on the system sizes and resetting rates. However, we only consider the power-law region in our paper, therefore, in Fig. \ref{fig_msd} (a), we only plot the results without very large times.
  
To further characterize the crossover behavior, the diffusion exponent is calculated as  

\begin{equation}
\alpha(r,L)\equiv \langle \frac{\partial \langle \Delta M^2\rangle}{\partial t}\rangle,
\end{equation}
the bracket $'\langle \rangle'$ outside $\frac{\partial \langle \Delta M^2\rangle}{\partial t}$ represents the averages over multiple independent samples.

In Fig. \ref{fig_msd} (b), the solid line denotes $r_c\sim L^{-z}$. It means that for the Ising model with a specific system size, if $r\lesssim L^{-z}$, then the stochastic resetting will not change the critical dynamical behavior of the order parameter. Similar results were also found for the probability distribution as explained in the last subsection (Sec. \ref{sec3.1}). 

Besides, for those black dots shown in Fig. \ref{fig_msd} (b), most of them denote the values of the diffusion exponents $\alpha(r, L)$ that is smaller than $\gamma/(\nu z)$. It is because we study the finite size systems ($L\lesssim 256$), and $\gamma/(\nu z)$ is only expected for system sizes approach infinity \cite{zhong2018generalized}.
 
\subsection{Measurement of the correlation time}

To further understand the crossover behavior of the MSD of $m$, we calculate the auto-correlation function of $m$ as
\begin{equation}
C(t) = \langle m(t) m(0)\rangle.
\end{equation}

From which we measure the correlation time by fitting the data of $C(t)$ as
\begin{equation}
C(t)\sim \exp(-t/ \tau).
\label{autocorr_m}
\end{equation} 

\begin{figure}[htb]
  \includegraphics[width=0.93 \linewidth]{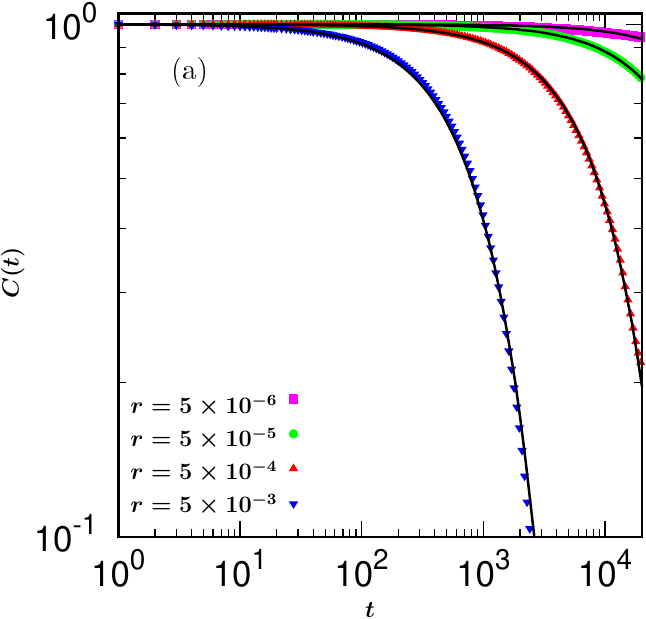}\hspace{3mm}
  \includegraphics[width=0.95 \linewidth]{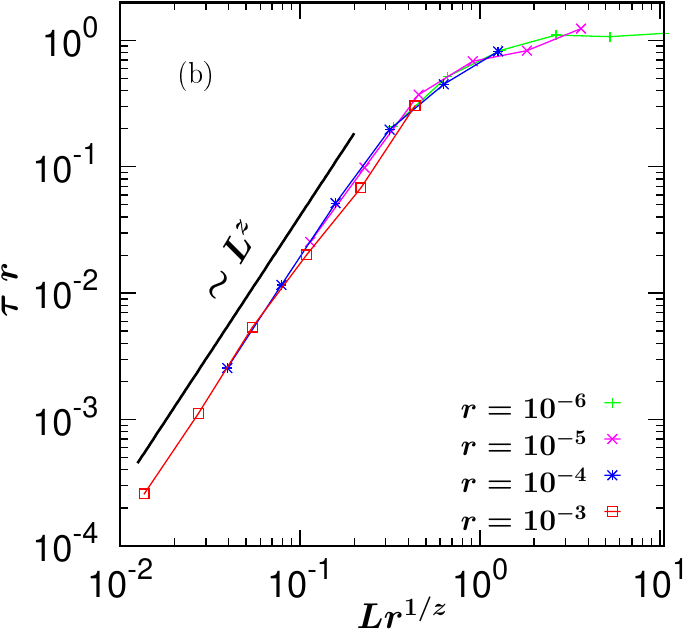}
  \caption{(a) The autocorrelation function of the magnetization $C(t)$ for $L=256$ with different resetting rate. The solid lines represent the best fit of the expression of Eq. (\ref{autocorr_m}), which provides the values of correlation time $\tau$. (b) The correlation time for different system sizes and resetting rates. The data collapse indicates that for $r \lesssim L^{-z}$, we have $\tau \sim L^z$. However, if $r$ is large, then the correlation time approaches a constant value of $1/r$. \label{fig_tau}}
\end{figure}

As an example shown in Fig. \ref{fig_tau} (a), for different resetting rates, the Eq. (\ref{autocorr_m}) (the solid lines) fits the simulation results of the autocorrelation function $C(t)$ well, which provide us the values of the correlation time $\tau$ for different resetting rates. 

Next, we plot the correlation time in Fig. \ref{fig_tau} (b). It explains that when $Lr^{1/z} \lesssim 1$, {\it i.e.}, $r \lesssim L^{-z}$, then $\tau \sim L^z$. When $r$ becomes larger, the correlation time saturates and it is not $L$ dependent anymore. It is only dependent on the resetting rate as $\tau \sim r^{-1}$.  

In summary, our observations with the $2D$ Ising model at the critical point reveal the presence of a characteristic resetting rate $r_c \sim L^{-z}$. Below $r_c$, stochastic resetting does not alter the system's dynamical properties. Additionally, as $L\rightarrow \infty$, $r_c \rightarrow 0$, indicating that in the thermodynamic limit, stochastic resetting does not impact the system's dynamics.

\begin{figure}[htb]
\includegraphics[width=0.95 \linewidth]{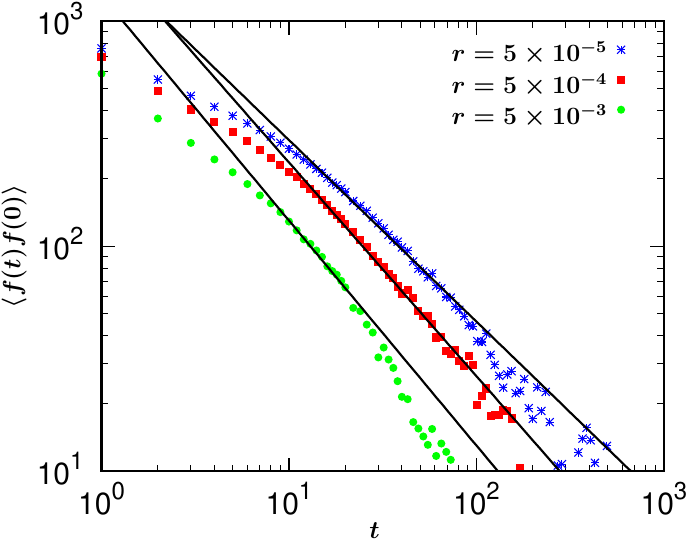} 
\caption{The autocorrelation function of the "restoring force" for different resetting rates and $L=256$. The solid lines are the power-law fit via Eq. (\ref{forceauto}).}
\label{fauto}
\end{figure}

\section{Conclusion \label{sec4}}
 
 We study the 2D Ising model with stochastic resetting at its critical point. Our results identify a threshold value of the stochastic resetting rate $r_c \sim L^{-z}$. Below this threshold, stochastic resetting does not influence critical dynamical properties of the order parameter. Notably, when $r>r_c$, we observe a crossover in the order parameter's behavior from anomalous to normal diffusion as the resetting rate increases. Since we have recognized that the subdiffusion of the order parameter in the Ising model likely follows the fBm type \cite{zhong2018generalized}, the results in this paper offer valuable insights into how stochastic resetting affects systems with anomalous diffusion of the fBm type.

 \section*{Appendix: Autocorrelation function of the "restoring force"}

 In Ref. \cite{zhong2018generalized}, it was argued that in the Ising-like systems, the observed anomalous diffusion of the order parameter at $T_c$ probability belongs to the fractal Brownian motion type. To confirm this argument, we calculate the autocorrelation function of the "restoring force" $f(t)$, which is the force that resists the change of the target magnetization. 
 
 Following the procedure suggested by Ref. \cite{zhong2018generalized}, assuming the system is thermalized, then we begin to fix the value of the magnetization. Note that when saying fix the magnetization does not mean the whole system is frozen, the spins can still evolve via the non-local Kawasaki spin-exchange dynamics\cite{newman1999}, {\it i.e.},  at each time step, we randomly select two spins $i$ and $j$ and we try to exchange the spin values by the Metropolis flip rule. Then at each regular time interval, the restoring force is measured as
 \begin{equation}
 f(t)=\sum_{i\in tagged} (-2 s_i) Min(1,e^{\Delta E_i/(k_B T_c)})
 \end{equation}
 
 Finally, we calculate the autocorrelation of the restoring force by $\langle f(t)f(0)\rangle$. As suggested by Ref. \cite{zhong2018generalized}, we should have
 \begin{equation}
 \langle f(t) f(0)\rangle \sim t^{-\alpha(r,L)},
 \label{forceauto}
 \end{equation}
 where $\alpha(r, L)$ is the diffusion exponent, and its value is adopted from Sec. \ref{sec3.2}.
 
The numerical results shown in Fig. \ref{fauto} demonstrate that the force autocorrelation function indeed follows the behavior of Eq. (\ref{forceauto}). It indicates that although the stochastic resetting changes the diffusion behavior with increasing diffusion exponents when $r$ increases, the anomalous diffusion for different resetting rates still belongs to the fBm type.

\section*{Acknowlegement \label{sec5}}

We would like to thank the helpful discussion with Y.J. Deng. Y.S.Chen is supported by the Natural Science Foundation of Fujian Province, China (Grant No. 2021J011027), and the start-up grant of Minjiang University. W.Z. acknowledges support from the National Natural Science Foundation of China Youth Fund (Grant No. 12105133), the Natural Science Foundation of Fujian Province, China (Grant No. 2021J011030, 2023J02032), and the start-up grant of Minjiang University (No.MJY21040).

\end{document}